\documentclass{elsart1p}
\usepackage{graphicx}
\usepackage{graphicx}
\usepackage{graphics}
\usepackage{amsmath}
\usepackage{epsfig}
\usepackage{epsf}
\newcommand {\sla}[1]{ #1 \!\!\!/}
\usepackage{amssymb}
\begin{document}
\begin{frontmatter}
%
%
%
\title{The Two-Boson-Exchange Correction to Parity-Violating Elastic Electron-Proton Scattering}
%
%
\author{Shin Nan Yang$^1$,
  Keitaro Nagata$^2$,
  Hai Qing Zhou$^{1,3}$,
 Chung Wen Kao$^2$}
\address{$^1$Department of Physics and  Center for Theoretical Sciences,
National Taiwan University,\\
Taipei 10617, Taiwan\\  $^2$Department of Physics, Chung-Yuan
Christian University, Chung-Li 32023, Taiwan\\
$^3$Department of Physics, Southeast University, Nanjing, China}

\begin{abstract}

 We calculate the two-boson-exchange (TBE) corrections to the parity-violating
asymmetry of the elastic electron-proton scattering in a simple
hadronic model including the nucleon and the $\Delta(1232)$
intermediate states. We find that  $\Delta$ contribution
$\delta_\Delta$ is, in general, comparable with the nucleon
contribution $\delta_N$ and the current experimental measurements
of strange-quark effects in the proton neutral weak current. The
total TBE  corrections to the current extracted values of
 $G^{s}_{E}+\beta G^{s}_{M}$ in recent experiments are found to lie
 in the range of $-7\sim +7\%$.

\end{abstract}
\end{frontmatter}
%
%
%
%

Strangeness content in the proton remains one of the most
intriguing questions in hadron structure. The parity-violating
asymmetry $A_{PV}=(\sigma_R-\sigma_L)/(\sigma_R+\sigma_L)$ in
polarized electron elastic scattering provides a rather clean
technique to probe the strangeness in the nucleon. Four
experimental programs SAMPLE , HAPPEX, A4, and G0 \cite{expts}
have been designed to measure this important quantity, which is
small and ranges from 0.1 to 100 ppm. This calls for greater
efforts to reduce theoretical uncertainty in order to arrive at a
more reliable interpretation of experiments.

The radiative corrections to $A_{PV}$ have been discussed in
\cite{Marciano}. However, theoretical uncertainties remain.
Recently, the contribution of the interference of the
two-photon-exchange (TPE) process of Fig. 1(a) with Born diagrams
has been evaluated in \cite{Afanasev05} in a parton model using
GPDs. It was found that indeed the TPE correction to $A_{PV}$ can
reach several percent in certain kinematics, becoming comparable in
size with existing experimental measurements of strange-quark
effects in the proton neutral weak current. However, the partonic
calculations of \cite{Afanasev05} are reliable only for $Q^2$ large
comparable to a typical hadronic scale, while all current
experiments  have been performed at lower $Q^2$ values. In addition,
the $\gamma Z$-exchange diagram of Fig. 1(b), expected to be of the
same order as the TPE correction, was not considered in
\cite{Afanasev05}.

\begin{figure}[h]
\begin{minipage}[h]{60mm}
\hspace{0.1cm} \noindent In this contribution, we report our recent
calculations \cite{zhou07,nagata08} of the TPE and $\gamma
Z$-exchange corrections to $A_{PV}$  with $N$ and $\Delta(1232)$
intermediate states as shown in  Figs. 1(a) and 1(b), where both are
treated in the same hadronic model developed in \cite{Blunden03}.

~At hadron level, the couplings of photon and $Z$-boson with proton
are well-known, while the couplings with the $\Delta$ are given as
\cite{nagata08},
\end{minipage}
\hspace{\fill}
\begin{minipage}[h]{70mm}
\centerline{\epsfxsize 1.4 truein\epsfbox{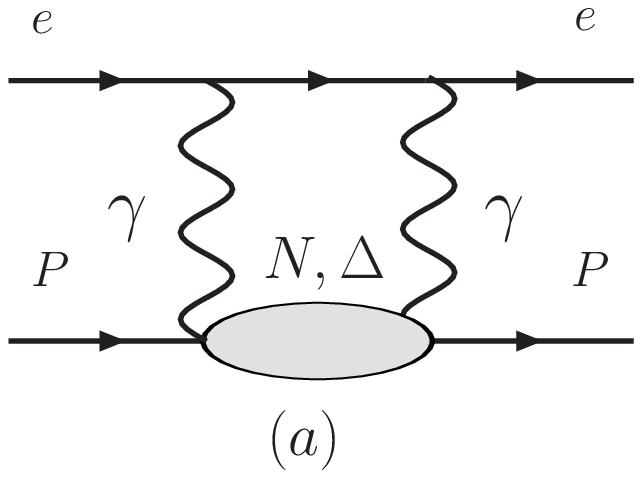}
\epsfxsize 1.34 truein\epsfbox{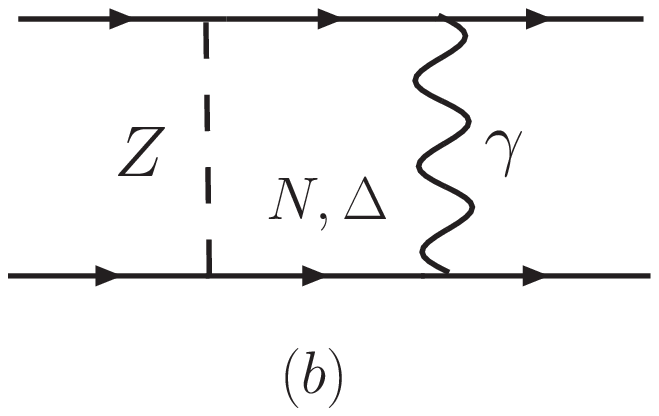}} \caption{(a)
TPE, and (b) $\gamma Z$-exchange diagrams for elastic
electron-proton scattering. Corresponding cross-box diagrams are
implied.}
\end{minipage}
\end{figure}
\begin{eqnarray}
&&\langle N(p')|J^{\gamma}_\mu|\Delta (p)\rangle =\frac{
F_{\Delta}(q^2)}{M_{N}^{2}}\overline{u}(p')
[g_{1}(g^{\alpha}_{\mu}\sla{p}\sla{q} -p_{\mu}\gamma^{\alpha}\sla{q}
-\gamma_{\mu}\gamma^{\alpha}p\cdot q+\gamma_{\mu}\sla{p}q^{\alpha})\nonumber\\
&&~~+g_{2}(p_{\mu}q^{\alpha}-p\cdot q g^{\alpha}_{\mu}) +g_{3}/M_{N}
(q^2(p_{\mu}\gamma^{\alpha}-g^{\alpha}_{\mu}\sla{p})
+q_{\mu}(q^{\alpha}\sla{p}-\gamma^{\alpha}p\cdot
q))]\gamma_{5}T_3u_{\alpha}^{\Delta}(p),\nonumber\\
&&\langle p(p')|J^{Z,V}_\mu|\Delta^{+}(p)\rangle
=\frac{F_{\Delta}(q^2)}{M_{N}^{2}}\overline{u}(p')
[\tilde{g}_{1}(g^{\alpha}_{\mu}\sla{p}\sla{q}
-p_{\mu}\gamma^{\alpha}\sla{q}-\gamma_{\mu}\gamma^{\alpha}p\cdot
q+\gamma_{\mu}\sla{p}q^{\alpha})\nonumber \\
&&~~+\tilde{g}_{2}(p_{\mu}q^{\alpha}-p\cdot q g^{\alpha}_{\mu})
+\tilde{g}_{3}/M_{N}(q^2(p_{\mu}\gamma^{\alpha}-g^{\alpha}_{\mu}\sla{p})
+q_{\mu}(q^{\alpha}\sla{p}-\gamma^{\alpha}p\cdot q))]\gamma_{5}u^\Delta_{\alpha}(p),\nonumber \\
&&\langle p(p')|J^{Z,A}_\mu|\Delta^{+}(p)\rangle = \frac{
H_{\Delta}(q^2)}{M_{N}^{2}}\overline{u}(p')
[h_{1}(g^{\alpha}_{\mu}(p\cdot
q)-p_{\mu}q^{\alpha})+h_{2}/M_{N}^{2}(q^{\alpha}q_{\mu}\sla{p}\sla{q}\nonumber \\
&&~~-(p\cdot q)\gamma^{\alpha}q_{\mu}\sla{q}) +h_{3}((p\cdot
q)\gamma^{\alpha}\gamma^{\mu}-\sla{p}\gamma_{\mu}q^{\alpha})
+h_{4}(g^{\alpha}_{\mu}p^{2}-p_{\mu}\gamma^{\alpha}\sla{p})]u^\Delta_{\alpha}(p),
\label{vertex}\end{eqnarray} where $M_N$ is the proton mass and
$q=p'-p$, $F's$ and $H's$ are the corresponding form factors. The
vertex functions in Eq. (\ref{vertex})  satisify constraints which
ensure only the physical spin-3/2 component of the Rarita-Schwinger
spinor couples with $\gamma$ and $Z$ \cite{nagata08}.

The form factors $F_\Delta$ and $H_\Delta$ are chosen to have
dipole form with a cut-off value of $0.84$ and $1.0 \,GeV$,
respectively. $\tilde{g}_{i}'s$ and $g_{i}'s$ are related by
$\tilde{g}_{i}=\sqrt{2/3}\, \alpha_{V} g_{i},$ with
$\alpha_V=(1-2\sin^2\theta_{W})/(2\cos\theta_{W}).$ Coupling
constants $g_i's$ are determined from $\gamma N\rightarrow \Delta$
at real photon point, while we infer the values of $h_i's$ from
the data of $\nu N\rightarrow \mu \Delta$ \cite{nagata08}.

We obtain the TPE and $\gamma Z$-exchange corrections to $A_{PV}$ as
shown in Fig. 2, where we plot $\delta$, defined by
\begin{equation}A_{PV}(1\gamma+Z+2\gamma+\gamma
Z)=A_{PV}(1\gamma+Z)(1+\delta_N+\delta_\Delta),\end{equation} {\it
vs.} $\epsilon \equiv [1+2(1+\tau)\tan^2\theta_{Lab}/2]^{-1}$, at
two different values of $Q^2= 0.1,\,1.5\,\,GeV^2$. $\theta_{Lab}$ is
the laboratory scattering angle and $\tau=Q^2/4M^2$. The subscripts
N and $\Delta$ in $\delta$ refer to effect associated with N and
$\Delta$ intermediate states, respectively.

\begin{figure}[h,b,t]
\centerline{\epsfxsize 4.0 truein\epsfbox{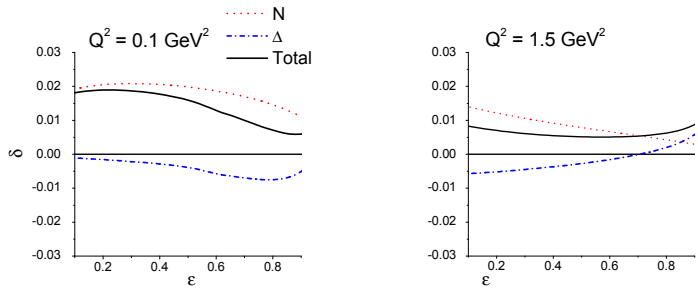}}
\caption{Two-boson-exchange corrections to parity-violating
asymmetry as functions of $\epsilon$ from 0.1 to 0.99 at $Q^2=0.1,\,
1.5 \,\,GeV^2$.} \label{Fig_delta}
\end{figure}

We see from Fig. \ref{Fig_delta} that $\delta$ is sensitive {\it
w.r.t.} both $\epsilon$ and $Q^2$ . For small $Q^2$,
$\delta_{\Delta}$ is almost zero and negative at small $\epsilon$,
and  decreases further as $\epsilon$ increases before turning around
at $\epsilon\sim 0.8$. These behaviors are in sharp contrast with
$\delta_N$ which always stays positive and decreases with increasing
$\epsilon$. Only at larger $\epsilon$, $\delta_\Delta$ is comparable
with $\delta_N$.

The strangeness form factors $G_E^s+\beta G_M^s$ extracted in the
current experiments have already taken into account of the $\gamma
Z-$exchange effects as estimated in \cite{Marciano} within a zero-

\begin{table}[hbt]
\begin{minipage}[h]{75mm}
\hspace{0.3cm}
\begin{tabular}
{|c|c|c|c|c|c|c|}
\hline  & I & II & III & IV & V &VI\\
\hline $Q^2(GeV^2)$&\,0.477\,  & \,0.109\,   &\, 0.23\,  &\,0.108 \,&\,0.262\,&\, 0.410\,\\
\hline $\epsilon$  & 0.974 & 0.994   & 0.83 & 0.83 &0.984&0.974 \\
\hline $\delta_{N}(\%)$  & 0.25& 0.34& 0.86  & 1.30 &0.28&0.275 \\
\hline $\delta_{\Delta}(\%)$ & 1.20& 1.90&-0.025 & -0.65 &1.33&1.14\\
\hline $\delta_G(\%)$ &+6.74 &-7.53& -4.26& -4.82& -1.70 &-0.10\\
\hline
\end{tabular}
\vspace{0.2cm} \caption{The corrections $\delta_G$ to $G_E^s+\beta
G_M^s$ for HAPPEX, A4, and G0 experiments. (I, II), (III, IV), and
(V, VI) refer to the HAPPEX,  A4,  and  G0 data, respectively.}
\label{tab1}
\end{minipage}
\hspace{\fill}
\begin{minipage}[h]{55mm}
\normalsize \noindent{momentum-transfer approximation scheme. To
avoid double counting, we need to first remove   the $\gamma
Z$-exchange effects estimated with \cite{Marciano} and use our
results instead. The resultant percentage change in  values of the
strange form factors defined as,
$\overline{G}_E^s+\beta\overline{G}_M^s=(G_E^s+\beta
G_M^s)(1+\delta_G),$ where
$\overline{G}_E^s+\beta\overline{G}_M^s$ denotes the values would
be extracted with our estimation of TBE effects, is given in Table
1. We see that even though $\delta_N$ and $\delta_\Delta$ are of
the same sign and  }
\end{minipage}
\end{table}

\noindent of comparable size, the $\Delta$ effects actually reduce
the difference between the results of \cite{Marciano} and the
effects associated with only elastic nucleon intermediate states
as found in \cite{zhou07}. The resultant corrections $\delta_G$ to
the current extracted values for the strangeness form factor
$G_E^s+\beta G_M^s$  are then only of a few percent. The large
effects we find for both the nucleon and  the $\Delta$ excitation
  brings up the question of the contributions of other higher
resonances.



%
\end{document}